\documentclass[11pt]{amsart}
\usepackage{amsmath}
\usepackage{amssymb}
\usepackage{epsfig}
\usepackage{a4wide}

\newcommand{\Z}{\ensuremath{\mathbb{Z}}}
\newcommand{\R}{\ensuremath{\mathbb{R}}}

\newcommand{\Q}{\ensuremath{\mathbb{Q}}}

\renewcommand{\S}{\ensuremath{{\mathcal S}}}

\newcommand{\C}{\ensuremath{{\mathcal C}}}
\newcommand{\T}{\ensuremath{{\mathcal T}}}

\newcommand{\aeq}{\Leftrightarrow} 
\newcommand{\deltxy}{\ensuremath{\delta^{(2)}}}
\newcommand{\deltz}{\ensuremath{\delta^{(1)}}}
\newcommand{\densx}{\ensuremath{\mbox{\rm dens}^{(1)}}}
\newcommand{\densy}{\ensuremath{\mbox{\rm dens}^{(2)}}}
\newcommand{\densz}{\ensuremath{\mbox{\rm dens}^{(3)}}}
\newcommand{\bs}{\boldsymbol}

\DeclareMathOperator{\dd}{d\!}
\DeclareMathOperator{\inn}{int}

\DeclareMathOperator{\conv}{conv}
\DeclareMathOperator{\cl}{cl}

\newtheorem{thm}{Theorem}[section]
\newtheorem{lemma}[thm]{Lemma}
\newtheorem{prop}[thm]{Proposition}
\newtheorem{defi}[thm]{Definition}

\begin{document}

\title{SCD Patterns Have Singular Diffraction}

\author{M. Baake}
\address{Fakult\"at f\"ur Mathematik, Universit\"at Bielefeld, 
Postfach 100131, 33501 Bielefeld,  Germany}
\email{mbaake@mathematik.uni-bielefeld.de}

\author{D. Frettl\"oh}
\address{Institut f\"ur Mathematik und Informatik, Universit\"at
  Greifswald, Jahnstr. 15a, 17487 Greifs\-wald,  Germany \newline
  From Aug. 2004 on: Fakult\"at f\"ur Mathematik, Universit\"at
  Bielefeld, 33501 Bielefeld, Germany}

\email{dirk.frettloeh@math.uni-bielefeld.de}

\begin{abstract} Among the many families of nonperiodic tilings known
so far, SCD tilings are still a bit mysterious. Here, we determine the
diffraction spectra of point sets derived from
SCD tilings and show that they have no
absolutely continuous part, that they have a uniformly discrete pure point part
on the axis $\R {\bs e}^{}_3$, and that they are otherwise supported on a
set of concentric cylinder surfaces around this axis. For SCD tilings with
additional properties, more detailed results are given. 
\end{abstract} 

\maketitle

\section{\bf The tilings} \label{intro}
After the discovery of families of tiles that permit only aperiodic tilings, 
the question arose whether there exists a single tile that permits {\em only}
aperiodic tilings by copies of itself (an {\em aperiodic prototile}). A first
example, which gives tilings in Euclidean 3--space, was found by 
{\sc  P.\ Schmitt} in 1988. It was elaborated later by {\sc J.H.\ Conway} 
and {\sc L.\ Danzer} (cf.\ \cite{d}). In particular, they modified {\sc
Schmitt's} prototile to a convex one. We refer to these tilings --- which 
will be described in this section --- as SCD tilings. 

A {\em tiling} in $\R^d$ is a collection of tiles 
$\{T^{}_n\}^{}_{n \ge 0}$ which covers $\R^d$ and contains no overlapping
tiles, i.e., $\inn(T^{}_k) \cap \inn(T^{}_n) = \varnothing$ for $k \ne n$.  A 
{\em  tile}  is a nonempty compact set $T \subset \R^d$ with the property that 
$\cl(\inn(T))=T$. A tiling $\T$ is called {\em aperiodic}, if $\T+{\bs x}=\T$
implies ${\bs x}={\bs 0}$.  

The SCD tilings are built from a single kind of tile --- a single 
{\em prototile} --- which we refer to as {\em SCD tile}. 
Essentially, the main idea is that the only possible tilings are of
the following 
form: The tiles can be put together to form layers,
which extend in two dimensions; these layers can be
stacked, but only in such a way that two consecutive layers are rotated
against each other by an angle which may be incommensurate to
$\pi$. Then, the symmetry groups of the resulting tilings may still be 
nontrivial, even infinite, but they contain no translation. 
To achieve this, we allow only directly congruent copies of the tiles,
but no mirror images (cf. Section \ref{remarks}). 

\begin{figure}[t]

\epsfig{file=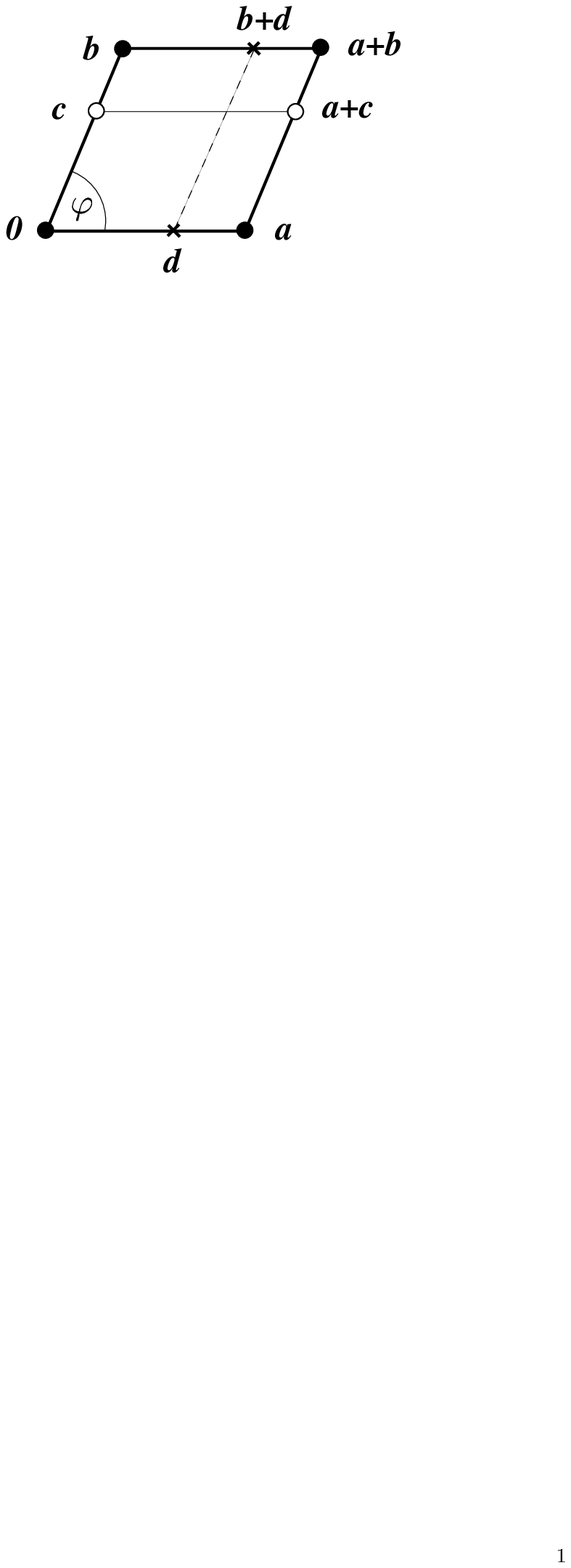} \hspace{20mm}
\epsfig{file=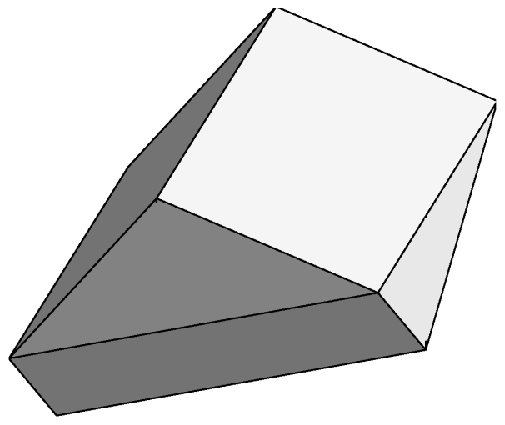} \label{scdtile}
\caption{The construction of an SCD tile (left) and a view of an SCD tile
  (right)}  
\end{figure}

\paragraph{\bf The SCD tile}
Choose $0 < \lambda < 1$, and positive real numbers $b_1,b_2,c$. 
Let $\varphi=\arctan(b_1/b_2)$, $a=\sqrt{b_1^2+b_2^2},$ 
and \[  {\bs a} =(a,0,0), \, {\bs  b}=(b^{}_1,b^{}_2,0),  
\, {\bs c}=\lambda {\bs b} +(0,0,c), \, {\bs d}=\lambda {\bs a} -
(0,0,c), \]  (cf.\ Fig.\ \ref{scdtile}).  Now, we define the SCD tile as  
\begin{equation} T=\conv ({\bs 0},{\bs a},{\bs b},{\bs a}+{\bs b},{\bs
c},{\bs a}+{\bs c},{\bs d},{\bs b}+{\bs d}), \label{tiledef} 
\end{equation} 
where $\conv(M)$ denotes the convex hull of $M$. The result is the union of
the two triangular prisms $\conv ({\bs 0},{\bs a},{\bs b},{\bs a}+\bs{b}
,{\bs c},{\bs a}+{\bs c})$ and $\conv 
({\bs 0},{\bs a},{\bs b},{\bs a}+{\bs b},{\bs d},{\bs b}+{\bs d})$, glued
together at the rhomb--shaped facet 
$\conv({\bs 0},{\bs a},{\bs b},{\bs a}+{\bs b})$. This is the reason that it 
is sometimes called Conway's biprism. If $\varphi\notin \pi  \Q$, we will 
call the SCD tile {\em incommensurate}  (which is the classical case),
otherwise {\em commensurate}.

We should mention that this is only one possible construction. 
Several generalizations or variations are possible (cf.\ Section \ref{remarks}
or \cite{d}). But all these tiles give rise to tilings with basically the same 
structure. 

\paragraph{\bf The SCD tilings}
Using translations of the SCD tile, one can put them together (i) by joining
triangular facets $\conv({\bs 0},{\bs b},{\bs c})$ with $\conv({\bs a},
{\bs a}+{\bs b},{\bs a}+{\bs c})$, and (ii) by joining triangular facets
$\conv({\bs 0},{\bs a},{\bs d})$ with  
$\conv({\bs b},{\bs b}+{\bs a},{\bs b}+{\bs d})$. If we do so inductively 
until no triangular facet remains uncovered, we end up with a planar
layer covering a 2--dimensional plane.  

This layer is congruent to $L=\{ {\bs x} + T \, | \, {\bs x} \in \varGamma \}$,
where $\varGamma$ is the 2--dimensional point lattice spanned by ${\bs a}$ and
${\bs b}$, i.e., $\varGamma = \Z {\bs a} + \Z {\bs b}$. The top of $L$ shows  
ridges and valleys, all parallel to each other, and all parallel to ${\bs b}$.
The bottom of $L$ also shows 'down under' (or upside down) valleys and ridges,
all of them parallel to ${\bs a}$. In order to stack the layers, consider a
layer  $L'=L-{\bs c}$. Take a second layer $L''=(0,0,c)+RL'$, where $R$ is a
rotation through $-\varphi$ around the axis $\R {\bs e}^{}_3=\langle (0,0,1)
\rangle_{\R}$.  $L''$ fits exactly on top of $L'$.  
In the same fashion, we proceed stacking layers and obtain
\begin{equation} \T = \bigcup_{m \in \Z} m(0,0,c) + R^m L', 
\label {tiling} \end{equation}
which is a tiling of $\R^3$. There are many other possibilities to build
SCD tilings. E.g., two consecutive layers can be shifted against each
other by an arbitrary translation in the direction of $R^m {\bs b}$, 
which is the direction of the matching valleys and ridges
of the two layers. Let us mention that {\sc Danzer}'s version
restricts these translations to a discrete set 
$\Z R^m {\bs  b}$ in order to allow crystallographic applications.
Therefore, 'SC tilings' might be a better notation for the more
general tilings we consider here. Nevertheless, we will stick to
the well--known
notation of SCD tilings throughout this paper, holding in mind that
the SCD tilings in \cite{d} are a proper subset of the SCD tilings
here.   
In this sense, all possible SCD tilings are congruent to 
\begin{equation}  \label{alltilings}
\T = \bigcup_{m \in \Z} m(0,0,c) + {\bs v}^{}_m + R^m L', 
\end{equation}
for some ${\bs v}^{}_m = (v^{(m)}_1,v^{(m)}_2,0)$, where ${\bs v}^{}_{m+1} - 
{\bs  v}^{}_m$  is a multiple of $R^m {\bs b}$. For a more thorough
discussion of all possible SCD tilings, see Section \ref{remarks} or
\cite{d}.  

As with tiles, we will distinguish between {\em incommensurate} SCD tilings,
if they are built from incommensurate SCD tiles, and {\em  commensurate} SCD
tilings otherwise. Now, it is easy to see 
that incommensurate SCD tilings are aperiodic: Since
$\varphi \notin \pi \Q$, all layers $m(0,0,c) + {\bs v}^{}_m + R^m L$
have pairwise different orientations. Consequently, a possible translation 
${\bs x}$ with $\T+ {\bs  x}=\T$ must map every layer onto itself. 
The translation vectors that fix the $m$--th layer $m(0,0,c) +
 {\bs v}^{}_m + R^m L$ are those in $R^m \varGamma$. So, the translation
vectors which fix the whole tiling are elements of 
\[ \bigcap_{m \in \Z} R^m \varGamma = \{ 0 \}, \]
wherefore all incommensurate SCD tilings are aperiodic. 

Note that even in the incommensurate case {\em finitely\/} many layers could
still possess nontrivial 
translation symmetries, as two layers might still share a so-called
coincidence site lattice $\varGamma'$ of finite index in
$\varGamma$. Then, any finite number of layers still may admit one,
where the index grows with the number of layers. In the limit of
infinitely many layers, only the trivial translation survives,
hence the final incommensurate SCD tiling is aperiodic, compare \cite{d} for 
details.

\section{\bf The diffraction spectrum}
Since the discovery of quasicrystals, a central point in the study of tilings
is the diffraction behaviour of tilings or point sets  (cf.\ \cite{so}). 
With point sets, one can model the structure of quasicrystals quite well, e.g.,
by representing every atom by a point. But many interesting structures were
originally described in terms of tilings. The usual way to examine the
diffraction behaviour of such structures is to replace every tile by one (or
more) reference points, in a way that the tiling and the point set determine
each other uniquely by local rules (i.e., they are 'mutually locally
derivable', cf.\ \cite{MB,BSJ}), and then to determine
the diffraction behaviour of the resulting point set. 
In this sense, {\em crystallographic} tilings in $\R^d$ --- i.e., tilings
which permit $d$ linearly independent translations --- correspond to 
crystallographic point sets, which again model ideal crystals. 
These show a sharp diffraction spectrum consisting of bright spots
only, the 'Bragg 
peaks', located on a uniformly discrete point set, compare \cite{Cowley,h}. 

The Fourier transform of structures like tilings or point sets (this will be
made precise below) gives a desription of their diffraction behaviour. E.g.,
the diffraction spectrum of quasiperiodic point sets, corresponding to 
physical quasicrystals (cf.\ \cite{h}), consists of Bragg 
peaks only, but their positions need not be discrete. In general,
any diffraction spectrum, described in terms of a positive measure $\mu$,
consists of three (unique) parts:  
\[ \mu = \mu^{}_{pp} + \mu^{}_{sc} + \mu^{}_{ac}, \]
compare \cite{bh} for examples and further references.
The {\em pure point} part $\mu^{}_{pp}=\sum_{{\bs x} \in \varLambda}
I({\bs x}) \delta^{}_{\bs x}$ is the sum of weighted Dirac measures 
(the so-called Bragg peaks) over a countable set $\varLambda$, where
$\delta^{}_{\bs x}$ is the normalized point measure at $\bs x$
(i.e., $\delta^{}_{\bs  x}(M)=1$, if ${\bs  x} \in M$, and
$\delta^{}_{\bs  x}(M)=0$ otherwise) and $I({\bs x})$ denotes the intensity.
The {\em singular continuous} part $\mu^{}_{sc}$ satisfies $\mu^{}_{sc}(\{
{\bs x} \}) = 0$ for all ${\bs x}$, but is supported (or concentrated) on a 
set of Lebesgue measure zero.  
The {\em absolutely continuous} part $\mu^{}_{sc}$ corresponds to a measure 
with a locally integrable density function and is supported on a set of 
positive Lebesgue measure.   
The diffraction spectrum $\mu$ of a structure is called {\em singular}, if
$\mu^{}_{ac}$ vanishes. It is called {\em pure point}, if $\mu^{}_{ac}$ and
$\mu^{}_{sc}$ vanish; i.e., if it consists of Bragg peaks only.  
The latter case occurs if the considered structure is a {\em model
  set} (cf. \cite{mo}). In this case, there is a rich theory one may
use to examine the diffraction spectrum. In this paper, however, we leave the
realm of pure point diffractive structures and have to use different
methods.   

This section makes use of the calculus of tempered distributions, also
known as generalized functions (compare \cite{rud,bb}). In particular, this
allows for a unified treatment of functions and measures. 
The following common notations are used:  
$\S(\R^d)$ denotes the Schwartz space of rapidly decreasing functions on
$\R^d$. The function  $\tilde{f}$ is given by $\tilde{f}(x)=\overline{f(-x)}$. 
The Fourier transform of $f$ is denoted by $\widehat{f}$. The tempered distributions,
$\S '$, are the continuous linear functionals on $\S(\R^d)$. For $T\in\S '$ and
$\varphi\in\S(\R^d)$, we will often write $\langle T,\varphi\rangle$ instead
of $T(\varphi)$.

As described above, one now constructs an {\em SCD set} $\varLambda^{}_{\rm SCD}$
from an SCD tiling and determines the diffraction spectrum of
$\varLambda^{}_{\rm SCD}$, taking up and extending previous work in this direction
\cite{d,n}. To do so, choose a point ${\bs z}$ in the interior of
the SCD tile $T$ in (\ref{tiledef}), choose an SCD tiling $\T$  and set
\[ \varLambda:=\varLambda^{}_{\rm SCD}:= 
   \{ {\bs v} +R^m {\bs z} \, | \, ({\bs v}+ R^m T) \in \T \}, \] 
i.e., replace every tile ${\bs v}+R^m T$ by the corresponding reference point 
${\bs v}+ R^m {\bs z}$. Obviously, $\varLambda$ consists of layers which 
are congruent to the lattice $\varGamma$. Now, define the measure 
\[ \omega := \omega^{}_{\rm SCD} := \sum^{}_{{\bs x} \in \varLambda}
    \delta^{}_{\bs x}. \] 
Let $C^{}_r=[-r/2,r/2]^3$ be the closed cube of sidelength $r$ centered at the
origin.  The diffraction spectrum of $\varLambda$ is described by the Fourier
transform $\widehat{\gamma}$ of the autocorrelation
\[ \gamma = \lim_{r \to \infty} r^{-3} \sum_{{\bs x},{\bs y} \in \varLambda
  \cap   C^{}_r} \delta^{}_{{\bs x}-{\bs y}}, \]
where the limit of these measures is taken in the vague topology. A priori,
it is not clear whether this limit exists. But since the considered measures 
are translation bounded, there is at least one convergent subsequence
\cite[Prop.\ 2.2]{h}. In this case, we go over to this convergent
subsequence. If  there is more than one convergent subsequence, we consider 
each one separately. This way, we can now always assume that
$\gamma$ exists as a tempered measure.  

Let $\omega^{}_r = \sum_{{\bs x} \in \varLambda \cap C^{}_r} 
\delta^{}_{\bs  x}$.  Then,
\[ \gamma = \lim_{r \to \infty} r^{-3} \omega^{}_r \ast
   \tilde{\omega}^{}_r \] 
where $\tilde{\omega}^{}_r := (\omega^{}_r)^{\tilde{}}$.
By definition, this means  that     
$ \lim_{r \to \infty} r^{-3} \langle \omega^{}_r \ast \tilde{\omega}^{}_r ,
\varphi \rangle$  exists for all test functions $\varphi \in \S(\R^3)$. So, 
\begin{eqnarray*}
\lim_{r \to \infty} r^{-3} \langle \omega^{}_r \ast \tilde{\omega}^{}_r
, \varphi \rangle & = & \lim_{r \to \infty} r^{-3} \int_{\R^3} \int_{\R^3}
\varphi({\bs x}+{\bs y})  \dd \tilde{\omega}^{}_r({\bs y}) \dd
\omega^{}_r({\bs x}) \\ 
 & = & \lim_{r \to \infty} r^{-3} \int_{C^{}_r} \int_{\R^3} \varphi({\bs x}+
{\bs  y})  \dd \tilde{\omega}^{}_r({\bs y}) \dd \omega({\bs x}) \\
& = & \lim_{r \to \infty} r^{-3} \int_{\R^3} \int_{\R^3} \varphi({\bs x}+
{\bs  y})  \dd \tilde{\omega}^{}_r({\bs y}) \dd \omega({\bs x}) \\[2mm]
& = & \lim_{r \to \infty} r^{-3} \langle \omega \ast
 \tilde{\omega}^{}_r, \varphi \rangle 
\end{eqnarray*}
and therefore 
\begin{equation} 
\gamma = \lim_{r \to \infty} r^{-3} \omega \ast \tilde{\omega}^{}_r .
\end{equation}
This can also be deduced from Lemma 1.2 in \cite{sch}.

In order to determine $\widehat{\gamma}$, we compute the Fourier transform 
of $\lim_{r \to \infty} r^{-3} \omega \ast \tilde{\omega}^{}_r$. 
Since the Fourier transform is continuous on the set $\S'$ of tempered
distributions, we have
\[ (\lim_{r \to \infty} r^{-3} \omega \ast \tilde{\omega}^{}_r 
   )^{\widehat{}}\, = \lim_{r \to \infty} r^{-3} (\omega \ast
   \tilde{\omega}^{}_r    )^{\widehat{}}. \] 
So, we proceed to compute $(\omega \ast \tilde{\omega}^{}_r )^{\widehat{}}\,$. 
Since $\tilde{\omega}^{}_r$ has compact support,
we have $\widehat{\tilde{\omega}^{}_r} \in \C^{\infty}$ and
$\tilde{\omega}^{}_r \ast \omega = \omega \ast \tilde{\omega}^{}_r$. The 
convolution theorem for distributions yields 
\begin{equation}  \langle (\omega \ast \tilde{\omega}^{}_r)^{\widehat{}}\,
  , \varphi \rangle = \langle \widehat{\omega},
  \widehat{\tilde{\omega}^{}_r} \cdot \varphi \rangle \label{faltsatz}
\end{equation}
for all $\varphi \in \S(\R^3)$. Let us take a closer look at $\omega$. 
It can be written as 
\[ \omega = \sum_{m \in \Z} \deltxy_{{\bs v}^{}_m + R^m \varGamma} \otimes
\deltz_{mc}, \]
where ${\bs v}^{}_m=(v^{(m)}_1,v^{(m)}_2)$, compare (\ref{alltilings}). Here
and in what follows, $\delta^{}_M := \sum_{{\bs x} \in M}  \delta^{}_{\bs x}$.
Note that $\deltxy_{{\bs v}^{}_m + R^m \varGamma}$ is a measure on $\R^2$
and $\deltz_{mc}$ is one on $\R^1$. Let $\varphi \in \S(\R^3)$ be of the form
$\varphi(x^{}_1,x^{}_2,x^{}_3) = f( x^{}_1,x^{}_2) g(x^{}_3)$, i.e., $f \in
\S(\R^2), g \in \S(\R)$, and $\varphi = f \cdot g$. Since linear
combinations of such functions $\varphi$ are dense in 
$\S(\R^3)$, the following calculation for tempered distributions holds,
\[ \widehat{\omega} = \sum_{m \in \Z} \widehat{\deltxy_{{\bs v}^{}_m + R^m
    \varGamma}}  \otimes \widehat{\deltz_{mc}}. \]
It remains to examine $\widehat{\deltz_{mc}}$, which equals 
$e^{-2\pi imc  x_3}$,  and
\begin{equation*}
\begin{split} 
\widehat{\deltxy_{{\bs v}^{}_m + R^m \varGamma}} & = 
  (\deltxy_{{\bs v}^{}_m} \ast   \deltxy_{R^m    \varGamma}
  )^{\widehat{} }\, 
  =  \widehat{\deltxy_{{\bs v}^{}_m}} \cdot \widehat{\deltxy_{R^m
  \varGamma}}  \\ 
& =  e^{-2\pi i (x^{}_1 v_1^{(m)} + x^{}_2  v_2^{(m)})} \densy(\varGamma)
\deltxy_{R^m  \varGamma^{\ast}},    
\end{split}
\end{equation*}
where $\densy$ denotes the 2--dimensional density of $\varGamma$.  
The last equality uses the Poisson summation formula in distribution form
\cite[p.~254]{Sch}
\begin{equation} \label{psf} 
\widehat{\delta^{}_{\varGamma}}=
{\rm dens}(\varGamma)\delta^{}_{\varGamma^{\ast}},   
\end{equation}  
where $\varGamma^{\ast} = \{ {\bs y} \, | \, {\bs y}{\bs x} \in \Z \;
\mbox{for all} \;  {\bs x} \in \varGamma \}$ denotes the dual (or reciprocal) 
lattice. The dual lattice of $R^m \varGamma$ is indeed $R^m \varGamma^{\ast}$,
since 
\[ \begin{array}{rcccrcl}
{\bs y} \in (R^m \varGamma)^{\ast} & \aeq & \forall {\bs x}' \in R^m
\varGamma:  \, {\bs y}{\bs x}' \in \Z & \aeq & \forall {\bs x} \in \varGamma:
\, {\bs y}R^m{\bs x} \in \Z & &  \\ 
& \aeq & \forall {\bs x} \in \varGamma: \, R^{-m}{\bs y}{\bs x} \in \Z  
& \aeq & R^{-m}{\bs y} \in \varGamma^{\ast} & \aeq & {\bs y} \in R^m
\varGamma^{\ast}  .   \end{array}  \]

Altogether, we get the following result.
Let $\varphi({\bs x}) = 0$ for all ${\bs x} \in M'=\left(
\bigcup_{m \in \Z} R^m \varGamma^{\ast} \right) \times \R$ (and
thus $\varphi({\bs x}) = 0$ for all  ${\bs x} \in  M:=\cl (M')$, since
$\varphi$ is continuous); in other words, let the support of $\varphi$ be
contained in the complement of $M$. Then,
\begin{equation} \label{ftomegatd}
\langle \widehat{\omega} , \varphi \rangle  =  \langle  \sum_{m \in \Z} 
e^{-2\pi i (x^{}_1 v_1^{(m)} + x^{}_2 v_2^{(m)}+mcx^{}_3)}
\densy(\varGamma) \deltxy_{R^m  \varGamma^{\ast}}  ,
\varphi \rangle = 0,
\end{equation}
where $\widehat{\omega}$ is already known to be a tempered  distribution.
Since the term $\deltxy_{R^m  \varGamma^{\ast}}$ refers only
to the two coordinates $x_1,x_2$, we conclude that 
the support of $\widehat{\omega}$ is a subset of $M$, as is 
the support of $(\tilde{\omega}^{}_r \ast \omega)^{\widehat{}}\,$, by
(\ref{faltsatz}).  So, the support of $\widehat{\gamma}$ is a subset of
$M$. So far, we have established: 
\begin{thm} \label{support-thm}
   The diffraction spectrum of any SCD set $\varLambda^{}_{\rm SCD}$ is 
   a singular measure, and
   it is supported  on the set \[ M = \cl \left(   \bigcup_{m \in \Z} R^m
  \varGamma^{\ast}\right) \times \R . \]  \hfill $\square$ 
\end{thm}
In the case of incommensurate SCD tilings, $M$ is the union of all concentric
cylinder surfaces $S$ with central axis $\R{\bs e}^{}_3$, where the 
radius of each $S$ is $\| {\bs  v} \|$ for some ${\bs v} \in \varGamma^{\ast}$.
In the case of commensurate SCD tilings, $M$ is a union of lines
parallel to $\R {\bs e}^{}_3$.  In this case, as we will see later on
in an example, the support of $\hat{\gamma}$ is a true subset of $M$. 

Now, take a closer look at the diffraction spectrum along 
$\R {\bs e}^{}_3$. From \eqref{ftomegatd}, we conclude 
\begin{equation} \label{ftomega}
\widehat{\omega} = \densy(\varGamma) \sum_{m \in \Z} e^{-2\pi i (x^{}_1   
v_1^{(m)} +x^{}_2 v_2^{(m)}+mcx^{}_3)} \deltxy_{R^m  \varGamma^{\ast}} ,
\end{equation} 
which might not be a measure in $\R^3$, but has a clear meaning as a
tempered distribution. 
The contribution to $\delta^{(2)}_0$ can be calculated by means of
\eqref{psf}  as follows,
\begin{equation} \label{psftrick} 
 \sum_{m \in \Z} e^{-2\pi i mc x^{}_3} = \sum_{n \in c\Z}\widehat{\deltz_n} = 
 (\deltz_{c\Z})\,\widehat{} = \densx(c\Z) \deltz_{(c\Z)^{\ast}} =   c^{-1}
  \deltz_{c^{-1} \Z}, 
\end{equation}
to be read as an equation for tempered distributions.

On the other hand, since $\tilde{\omega}^{}_r$ is a finite measure with
compact support, its Fourier transform is an analytic function and can
be written as
\begin{equation}  \label{ftomegar}
  \widehat{\tilde{\omega}^{}_r} ({\bs x}) = 
  \sum_{{\bs y} \in   \varLambda \cap C^{}_r} e^{2\pi i (x^{}_1 y^{}_1 + 
   x^{}_2 y^{}_2 + x^{}_3 y^{}_3)}.
\end{equation}
For ${\bs x} =(0,0,x^{}_3)$, we thus get 
\begin{equation*} \begin{split} 
\lim_{r \to \infty} r^{-3} \widehat{\tilde{\omega}^{}_r} \widehat{\omega}
  & =  \densy(\varGamma) \lim_{r \to \infty} r^{-3} \Biggl( \sum_{{\bs y} \in
  \varLambda \cap C^{}_r}   e^{2\pi i x^{}_3 y^{}_3} \Biggr) 
  \Biggl( \sum_{n \in \Z} 
  e^{-2\pi   i  nc x^{}_3} \;  \deltxy_{R^n  \varGamma^{\ast}} \Biggr) \\
& =  \densy(\varGamma) \lim_{r \to \infty} r^{-3}  \Biggl( \sum_{m=-\lceil r/2
  \rceil}^{\lfloor r/2   \rfloor} d_r^{(m)} r^2  e^{2\pi i mc y^{}_3 }  \Biggr)
  \Biggl(   \sum_{n \in \Z}   e^{-2\pi i nc x^{}_3} \deltxy_{R^n
  \varGamma^{\ast}}   \Biggr) 
\end{split} 
\end{equation*} 
Here, $d_r^{(m)}$ is chosen such that $d_r^{(m)} r^2$ counts the number of
elements of $\varLambda \cap C^{}_r$ in layer $m$. So, $d_r^{(m)}$ depends on
$\densy(\varGamma)$, and $\lim_{r \to \infty} d_r^{(m)}=
\densy(\varGamma)$ for all $m \in \Z$. 

Putting the pieces together, and restricting to the central axis, we obtain
\[  \lim_{r \to \infty} r^{-3} \widehat{\tilde{\omega}^{}_r} 
    \widehat{\omega} |^{}_{\R {\bs e}_3^{}} =
   c^{-1} (\densy(\varGamma))^2 \lim_{r \to \infty} r^{-1} \Biggl(
\sum_{m=-\lceil r/2   \rceil}^{\lfloor r/2   \rfloor}  e^{-2\pi i mc
  x^{}_3 }  \Biggr) \deltz_{c^{-1}\Z}. \]  
This expression vanishes for $x^{}_3 \notin c^{-1}\Z $, 
while for   $x^{}_3 \in   c^{-1}\Z$ we get
\[ c^{-1} (\densy(\varGamma))^2 \lim_{r \to \infty} r^{-1} 
\sum_{m=-\lceil r/2   \rceil}^{\lfloor r/2   \rfloor}  1  = c^{-1}
(\densy(\varGamma))^2 = \densy(\varGamma) \densz(\varLambda^{}_{\rm SCD}). \]
In analogy to $\densy$, $\densz$ denotes 3--dimensional density. It follows:
\begin{thm} The diffraction spectrum $\widehat{\gamma}$ of any SCD set
  $\varLambda^{}_{\rm SCD}$, restricted to   $\R {\bs e}^{}_3$, is
  pure point. In    particular,  
\[ \widehat{\gamma}|^{}_{\R{\bs e}^{}_3} = \densy(\varGamma) \,
\densz(\varLambda^{}_{\rm SCD}) \sum_{{\bs x} \in c^{-1} \Z {\bs e}^{}_3} 
\delta^{}_{\bs x}. \] \hfill $\square$
\end{thm}
For special cases, this result already appears in \cite{n}.
In the general case, it seems difficult to achieve results about the 
explicit behaviour on the cylinder surfaces. If the SCD tiling has additionial
properties, it is possible to show that all existing Bragg peaks are located on 
$\R {\bs e}^{}_3$. 

\begin{defi} 
A point set $\varLambda$ in $\R^d$ is called {\em repetitive}, 
if for every $r>0$ some $R > 0$ exists such that for all 
$x,y \in \R^d$ a congruent copy of $(x+C^{}_r) \cap \varLambda$ occurs in
every set  $(y + C^{}_R) \cap \varLambda$. 
\end{defi}
This definition has a natural extension to the repetitivity of tilings. For
our purposes, it suffices to call an SCD tiling repetitive, if the
corresponding SCD sets are repetitive.   
In particular, if $\T$ is repetitive, there are only finitely many ways how
two tiles can touch each other. (Otherwise, there would be infinitely many
different pairs of tiles, each fitting into a box $C_r$ with 
$r=2 \|{\bs a}+{\bs  b}\|$. This infinitely many pairs, having all the same
positive volume,  must be contained in a finite ball of radius $R$, which is
impossible.) 
\begin{prop} \label{repprop}
   If an SCD tiling $\T$ is repetitive, then $\varphi = \arccos(p/q)$, with
   $p,q \in \Z$. 
\end{prop}
\begin{proof} 
Let $\T$ be repetitive. Then the tiles of two consecutive layers $L^{}_i,
L^{}_{i-1}$ can touch each other in only finitely many ways. W.l.o.g., let
$L^{}_i= T + \varGamma$, $L^{}_{i-1}=R^{-1}(T+\varGamma) -R^{-1}{\bs c}$ and
$\varGamma = \langle (1,0),(b^{}_1,b^{}_2)\rangle_{\Z}$. By the definition of
$T$ and $\T$, it follows that $b_1=\cos(\varphi), b_2=\sin(\varphi)$ and 
that $R^{-1}$ (recall that $R$ is a rotation through the angle $-\varphi$) is
given by 
\[ \left( \begin{array}{cc} \cos(\varphi) & -\sin(\varphi) \\
\sin(\varphi) & \cos(\varphi) \end{array} \right) \]
So, $R^{-1}\varGamma = \langle (b^{}_1,b^{}_2),(b_1^2 - b_2^2, 2 b^{}_1 b^{}_2)
\rangle_{\Z}$. Obviously, $\langle (b^{}_1,b^{}_2) \rangle_{\Z} 
\subseteq \varGamma \cap R^{-1}\varGamma$. Since the tiles of
$L^{}_i$ and $L^{}_{i-1}$ touch each other in finitely many ways, there are
only finitely many possibilities, how a point of $\varGamma$ is positioned
relative to its nearest point in $R^{-1} \varGamma$. Consequently, one has
$(\varGamma \cap R^{-1}\varGamma) \setminus \langle (b^{}_1,b^{}_2)
\rangle_{\Z} \ne \varnothing$. Therefore, the equation 
\begin{equation} \label{coinceq}
 \kappa (1,0) + \lambda (b^{}_1,b^{}_2) = \mu (b^{}_1,b^{}_2) + \nu 
(b_1^2 - b_2^2, 2 b^{}_1 b^{}_2) 
\end{equation}
has a solution, where $\kappa \ne 0 \ne \nu$. We have to show that this
is only possible if $b_1$ is a rational number $p/q$. Let $b^{}_1$ be an
irrational number. From $\lambda b^{}_2 = \mu b^{}_2 - \nu 2 b^{}_1 
b^{}_2$ $(\lambda,\mu,\nu \in \Z)$, one concludes $\nu = 0$ and 
$\lambda = \mu$. Therefore, 
\[ \kappa  + \lambda b^{}_1  = \mu b^{}_1 + \nu (b_1^2 - b_2^2) \]
gives $\kappa=0$, so there is no solution of (\ref{coinceq}) with $\kappa \ne
0 \ne \nu$.   
\end{proof}

\begin{thm} \label{repthm}
Let $\varLambda$ be an incommensurate SCD set. If $R^m \varLambda + m{\bs c} =
\varLambda$ for some $m \ge 1$, or if $\varLambda$ is repetitive and
$\varphi=\arccos(p/q)$, where $q$ is odd, then the diffraction spectrum of
$\varLambda$  is singular continuous on $M \setminus \R {\bs e}^{}_3$.
\end{thm}

\begin{lemma} \label{taurlemma}
Let $R$ be an orthogonal map, $\mu$ a measure, and let the measure 
$R.\mu$ be given by $R. \mu (A) = \mu(R^{-1}A)$. Then
\[  R.\widehat{\mu} = \widehat{R. \mu}\, . \]
\end{lemma}
\begin{proof} Let $\varphi \in \S(\R^3)$. It is clear that
$\langle R. \mu, \varphi \rangle = \langle \mu, \varphi \circ R
  \rangle$. Since 
\[ \begin{array}{rclcl}
\widehat{\varphi \circ R}({\bs x}) 
& = & \int \varphi(R{\bs y}) e ^{-2 \pi i {\bs x}{\bs y}}  \dd {\bs y} 
& = & \int \varphi (\widetilde{\bs y}) e ^{-2 \pi i {\bs x}
 ( R^{-1}\widetilde{\bs y})} \dd \widetilde{\bs y} \\[1mm] 
& = & \int \varphi (\widetilde{\bs y}) e ^{-2 \pi i (R{\bs x})
  \widetilde{\bs y}} \dd \widetilde{\bs y} 
& = &  \widehat{\varphi}(R{\bs x}), \\
\end{array} \]
where $\widetilde{\bs y}=R{\bs y}$, it follows that $\widehat{\varphi \circ R} =
\widehat{\varphi} \circ R$. Thus
\[ \langle \widehat{R. \mu} , \varphi \rangle = \langle R.\mu ,
\widehat{\varphi} \rangle = \langle \mu , \widehat{\varphi} \circ R \rangle
=  \langle \mu , \widehat{\varphi \circ R} \rangle = \langle \widehat{\mu} ,
\varphi \circ R \rangle = \langle R.\widehat{\mu} , \varphi \rangle  \]
which proves the claim.
\end{proof}

\begin{proof}[Proof of Theorem~$\ref{repthm}$]
Let $R^m \varLambda + m {\bs c} = \varLambda$. The support of the
autocorrelation $\gamma$ is $\varLambda - \varLambda = \{ {\bs x} - {\bs y} \,
| \, {\bs x}, {\bs y} \in \varLambda \}$. Since 
\[ R^m(\varLambda - \varLambda) = R^m \varLambda + m{\bs c} - (R^m \varLambda
+  m {\bs  c}) = \varLambda - \varLambda, \] 
we get $\gamma = R^m.\gamma$. Lemma \ref{taurlemma} implies
$\widehat{\gamma}= \widehat{R^m.\gamma} = R^m.\widehat{\gamma}$, and therefore
$\widehat{\gamma} = R^{km}\widehat{\gamma}$ for all $k \in \Z$. 

Now, let $\varLambda$ be repetitive and $\varphi=\arccos(p/q)$, where $q$ is
odd. Like $\varLambda$ itself, the set $\varLambda - \varLambda$ consists of
equidistant layers. If $\varLambda = \bigcup_{k \in \Z} R^k \varGamma + 
{\bs  v}^{}_k +  k{\bs c}_0$ (where ${\bs c}_0=(0,0,c)$), then
\[ \varLambda - \varLambda = \bigcup_{i \in \Z} \bigcup_{k \in \Z}
R^{k+i} \varGamma + {\bs v}^{}_{k+i} + (k+i){\bs c}_0 - (R^k \varGamma + 
{\bs  v}^{}_k + k{\bs c}_0).  \]   
Now we use a fact from \cite{d}: If $\T$ is a repetitive SCD tiling, and if
$\varphi=\arccos(p/q)$, $q$ odd, then the union of $i$ consecutive  
layers in $\T$ is congruent to any other such union of $i$ consecutive
layers in $\T$. Therefore, all difference sets $R^{k+i} \varGamma + 
{\bs  v}^{}_{k+i} + (k+i){\bs c}_0 - (R^k \varGamma + {\bs v}^{}_k + 
k{\bs  c}_0)$ are congruent. This means ${\bs  v}^{}_{k+i} - {\bs  v}^{}_k=
R^k({\bs  v}^{}_{i} - {\bs  v}^{}_0)$. Since $R{\bs c}_0={\bs c}_0$, it
follows  
\begin{eqnarray*} 
R(\varLambda - \varLambda) & = & R \Biggl( \bigcup_{i \in \Z} 
\bigcup_{k \in \Z} R^k( R^i \varGamma - \varGamma + {\bs v}^{}_{i} - 
{\bs  v}^{}_0) + i{\bs c}_0 \Biggr) \\
& = & R \Biggl( \bigcup_{k \in \Z} R^k \biggl( \bigcup_{i \in \Z}  R^i
\varGamma  - \varGamma + {\bs v}^{}_{i} - {\bs v}^{}_0 + i{\bs c}_0 \biggr)
\Biggr) =  \varLambda - \varLambda. 
\end{eqnarray*}
Therefore, one has $\widehat{\gamma} = R^{k}.\widehat{\gamma}$ for
all $k \in \Z$.  

In both cases, the following argument applies: If there is 
a Bragg peak $I({\bs x}) \delta^{}_{\bs x}$ at ${\bs x} \in M \setminus 
\R  {\bs  e}^{}_3$ with intensity $I({\bs x})>0$, then there are infinitely
many Bragg peaks $I({\bs x}) \delta^{}_{R^{km} {\bs x}} \, (k \in \Z)$  
contained in a circle of diameter $\| {\bs x}\|$. But since $\widehat{\gamma}$
is a tempered distribution, it is bounded on every compact set $K \subset
\R^3$. This is a contradiction. Therefore, no Bragg peaks occur in $M
\setminus \R {\bs  e}^{}_3$.       
The claim now follows from Theorem~\ref{support-thm}.
\end{proof}

In contrast to this situation, let us ask
what happens for a fully periodic SCD tiling. This is only
possible if it is a commensurate SCD tiling (which means that  
$R$ is of finite order), and if the sequence $(v_1^{(m)},v_2^{(m)})$
is  periodic (to be precise: periodic mod $R^m {\bs a}$). Equivalently: There
is a $k \ge 1$, such that $R^k={\rm id}$ and $v_1^{(m+k)} \equiv v_1^{(m)},
v_2^{(m+k)} \equiv v_2^{(m)} \mod R^m {\bs a}$ for all $m \in \Z$. In this 
case, (\ref{ftomega}) gives 
\begin{eqnarray*} 
\widehat{\omega} & = & \densy(\varGamma) \, \sum_{n \in k\Z} \sum_{j=0}^{k-1}
  e^{-2\pi i (x^{}_1   v_1^{(j)} + x^{}_2   v_2^{(j)}+(n+j)cx^{}_3)}
  \deltxy_{R^{n+j}    \varGamma^{\ast}} \\
& = & \densy(\varGamma) \, \sum_{j=0}^{k-1} e^{-2\pi i (x^{}_1 v_1^{(j)} + x^{}_2
  v_2^{(j)}+jcx^{}_3)} \Biggl( \sum_{n   \in   ck\Z}  e^{-2 \pi i n x^{}_3}
  \Biggr)   \deltxy_{R^j  \varGamma^{\ast}} \\ 
& = & \densy(\varGamma) \, \sum_{j=0}^{k-1} e^{-2\pi i (x^{}_1 v_1^{(j)} + x^{}_2
  v_2^{(j)}+jcx^{}_3)}    (ck)^{-1} \bigl( \deltxy_{R^j \varGamma^{\ast}}
  \otimes    \deltz_{(ck)^{-1} \Z} \bigr)  \\ 
\end{eqnarray*}
This term vanishes everywhere except on $\left( \bigcup_{j=1}^k R^j
\varGamma^{\ast} \right) \times (ck)^{-1} \Z$. So, the diffraction spectrum of
a fully periodic SCD tiling is, as expected, supported on a uniformly
discrete point set.  It is, in fact, a pure point diffraction spectrum,
consisting of isolated Bragg peaks.
The support is indeed {\em uniformly} discrete, since
from the periodicity of the tiling the repetitivity follows, wherefore
Proposition \ref{repprop} yields $\varphi = \arccos(p/q)$ ($p,q \in \Z$). 
Since the tiling is commensurate, $(p,q)$ can take the values $(0,1)$ or
$(1,2)$ only.

\section{\bf Further remarks} \label{remarks} 
 
{\bf 1.} One special case which occurs is the body-centered cubic lattice
(bcc) as the underlying point set of an SCD tiling. It is the dual of the root
lattice $D^{}_3$, compare \cite{cs}: 
\[ {\rm bcc} = D_3^{\ast}= \langle  (1,0,0), (0,1,0),
(\tfrac{1}{2},\tfrac{1}{2},\tfrac{1}{2}) \rangle_{\Z}. \]   
This is achieved by placing the reference point ${\bs z}$ in the center
$\frac{1}{2}( {\bs a} + {\bs b})$ of the
SCD tile, and choosing (cf.\ Section \ref{intro}):
 \[ {\bs a}=(1,0,0), \, {\bs b}=(0,1,0), \, {\bs c}
=(0,\tfrac{1}{2},\tfrac{1}{2}), \, 
{\bs d}=(\tfrac{1}{2},0,-\tfrac{1}{2}),  \, v_1^{(m)}=v_2^{(m)}= 
 \left\{  \begin{array}{cl} 0\, , & m \; {\rm even} \\ \tfrac{1}{2}\, , & m
 \;  {\rm odd} \end{array} \right. \]
Using (\ref{ftomega}) and (\ref{ftomegar}), one finds for this case
\begin{equation*}  \widehat{\gamma^{}_{{\rm bcc}}} = \lim_{r \to \infty} r^{-3}
   \Biggl( \sum_{{\bs y} \in 
    {\rm bcc} \cap C^{}_r} e^{2\pi i {\bs x} {\bs y}} \Biggr) \Biggl(
  \sum_{m \in \Z} e^{-2\pi i (x^{}_1 v_1^{(m)} + x^{}_2 v_2^{(m)}+x^{}_3m/2)}
    \Biggr)  \densy(\Z^2)   \deltxy_{\Z^2}  
\end{equation*}
This term vanishes on $\{{\bs x} \, | \, (x^{}_1,x^{}_2) \notin \Z^2\}$. For
$(x^{}_1,x^{}_2) \in \Z^2$, one finds
\begin{eqnarray*}  & & \lim_{r \to \infty} r^{-3} \Biggl( \sum_{n = -\lceil r
    \rceil}^{\lfloor r \rfloor} r^2 e^{2\pi i x^{}_3 n/2} \Biggr) \Biggl(
  \sum_{m   \in 2\Z+1} e^{- \pi i (x^{}_1  + x^{}_2 +x^{}_3 m)} +
  \sum_{m \in 2\Z}   e^{-2\pi   i x^{}_3 m/2}  \Biggr) \\ 
& = & \lim_{r \to \infty} r^{-1} \Biggl( \sum_{n = -\lceil r \rceil}^{\lfloor r
  \rfloor} e^{2\pi i x^{}_3 n/2} \Biggr) \left( 1+ e^{- \pi i (x^{}_1  +
  x^{}_2 +x^{}_3)} \right) \sum_{m \in \Z}   e^{-2\pi i x^{}_3 m}.  
\end{eqnarray*} 
From (\ref{psftrick}), one gets $\sum_{m \in \Z}   e^{-2\pi i x^{}_3  m}
=\deltz_{\Z}$. So, this term vanishes for $x^{}_3   \notin \Z$, and for 
$x^{}_3 \in \Z$ we have to examine the factor $1+ e^{-\pi   i (x^{}_1  +
  x^{}_2 +x^{}_3)}$. It equals $2$ (resp.\ $0$) if 
$x^{}_1  + x^{}_2 +x^{}_3$ is even (resp.\ odd).
In the even case, the first sum does not converge, so the limit is not
zero. Altogether: The diffraction spectrum of bcc consists of Bragg peaks on
points in 
\[ D^{}_3 = \{ {\bs x} \, | \,   x^{}_1+x^{}_2+x^{}_3 \equiv 0 \mod 2 \}. \] 
In this way, we get the well--known result that the diffraction image of the
bcc is pure point, with Bragg peaks on the points of the dual lattice
$(D_3^{\ast})^{\ast}=D^{}_3=2\,{\rm fcc}$. 

In a similar way, one finds further structures that are well known from
crystallography or discrete geometry, such as the root lattices $\Z^3$ and
$D^{}_3$ (which is a scaled version of the face centered cubic lattice fcc), 
or the hexagonal close packing (\cite{cs}). 

{\bf 2.} The description of the SCD tile in Section \ref{intro} follows the
idea of {\sc Conway}. The prototile found by {\sc Schmitt} is
not convex, but showed itself the valleys and ridges, which occur on the
layers of our tilings (and his tilings have essentially the same structure as
ours). Anyway, both tiles lead to the same SCD sets, and both tiles are
examples of aperiodic prototiles.  But the latter is only 
true if we forbid tilings which contain both our SCD tile 
and its mirror image. E.g., let $T$ be as in (\ref{scdtile}) and $T'$ the
mirror image of $T$ under reflection in the plane spanned by ${\bs a}$ and 
${\bs  b}$. The layer $L=T + \varGamma$ contains only 
translations of $T$,  the layer $L'=T'+ {\bs c}+\varGamma$ contains only 
translations of $T'$. The tiling
\[ \T = \bigcup_{m \in 2\Z} (0,0,mc) + (L \cup L') \]
is invariant under the translations $t({\bs x})={\bs x}+(0,0,2 c)$ and 
$u({\bs  x})={\bs x}+{\bs b}$, hence not aperiodic.   

In our desription, the angle $\varphi$ can take any value in $]0, \pi/2[$. 
The SCD tile described by {\sc Danzer} uses $\varphi=\arccos(p/q)$, where
$p,q$ are positive integers, $p <q, \, q \ge 3$ (leading to 
incommensurate SCD tilings). In this case, it is possible to {\em enforce} 
SCD tilings which are repetitive. Then, in particular, two
tiles can touch each other only in finitely many different ways. (This is
clearly not true for all SCD tilings considered in this paper.) Using this,
one can modify the 
shape of the prototile in such a way that the occurrence of mirror images of
the prototile is ruled out. This can be done, e.g., by adding projections and
indentations to the tiles, fitting together like key and keyhole, but only
if the tiles are directly congruent. So, in this case, one has indeed a single
prototile  --- no longer convex --- permitting only aperiodic tilings, just by
its shape. 

Anyway, even in the last setting, there may occur other symmetries,
namely screw motions. Obviously, the tiling $\T$ in (\ref{tiling}) is
invariant under the map $s({\bs x})=R{\bs x}+(0,0,c)$. More generally, if we
choose an arbitrary SCD tile discussed here, then in the set of all 
tilings built from this tile we will always find
tilings invariant under the maps $s^k, \, (k \in \Z)$. Thus the symmetry
group of such tilings is infinite. 
In less than three dimensions, aperiodicity is equivalent to finiteness of the
symmetry group. The SCD tilings show that this is not true in
general. Therefore, it makes sense to rephrase the question 'Is 
there an aperiodic prototile?' as 'Is there a prototile that permits only
tilings with finite symmetry group?', shortly: 'Is there a {\em strongly
  aperiodic} prototile?' (cf.\ \cite{m}). To our knowledge, no
answer to this question is known so far.

{\bf 3.} To some extent, the underlying mechanism of SCD tilings does occur in
Nature. The structure of smectic $C^{\ast}$ liquid crystals resembles the
layer structure: planar, 2--periodic 'sheets' of tilted molecules (called
directors) are stacked with a screw order on top of each other \cite{g}. This
happens in such a way that the (effective) period in direction 
$\R {\bs e^{}_3}$ is on a much greater length scale than the elementary
periods within the layers.  

\subsection*{Acknowledgements}

The authors thank L.\ Danzer and K.-P.\ Nischke for helpful
discussions. This work was supported by DFG.

\end{document}